\documentclass[apj]{emulateapj}
 
\usepackage{amsmath}
\def\fermi{{\it Fermi\/}}
\def\swift{{\it Swift\/}}

\def\lsim{\mathrel{\lower .85ex\hbox{\rlap{$\sim$}\raise
.95ex\hbox{$<$} }}}
\def\gsim{\mathrel{\lower .80ex\hbox{\rlap{$\sim$}\raise
.90ex\hbox{$>$} }}}
\newbox\grsign \setbox\grsign=\hbox{$>$}
\newdimen\grdimen \grdimen=\ht\grsign
\newbox\laxbox \newbox\gaxbox
\setbox\gaxbox=\hbox{\raise.5ex\hbox{$>$}\llap
     {\lower.5ex\hbox{$\sim$}}}\ht1=\grdimen\dp1=0pt
\setbox\laxbox=\hbox{\raise.5ex\hbox{$<$}\llap
     {\lower.5ex\hbox{$\sim$}}}\ht2=\grdimen\dp2=0pt

\shorttitle{0FGL J1830.3+0617: A Fermi Blazar}
\shortauthors{Mirabal \& Halpern}

\begin{document}

\title{0FGL J1830.3+0617: A Fermi Blazar Near the Galactic Plane}

\author{N. Mirabal\altaffilmark{1,2}, \& J. P. Halpern\altaffilmark{3}}

\altaffiltext{1}{Ram\'on y Cajal Fellow; Dpto. de F\'isica At\'omica, 
Molecular y Nuclear, Universidad Complutense de 
Madrid, Spain}
\altaffiltext{2}{Email: mirabal@gae.ucm.es}
\altaffiltext{3}{Columbia Astrophysics Laboratory, Columbia University,
  550 West 120th Street, New York, NY~10027}

\begin{abstract}
We present a multiwavelength study of the 
unidentified \fermi\ $\gamma$-ray
source 0FGL J1830.3+0617, which
exhibits variability above 200 MeV on timescales of days to weeks. 
Within the \fermi\ 95\% confidence error contour lies
B1827+0617, a radio source with spectral index
$\alpha = 0.09$ between 1.4 and 4.85 GHz. 
The flat spectral index and flux density of 443 mJy
at 4.85 GHz are consistent with the bulk
of \fermi\ sources associated with blazars. 
It is also detected in the $0.3-10$ keV band by \swift.
Optical imaging in 2009 May identifies B1827+0617
at $R \approx 16.9$, and shows
that it is at least 2 magnitudes
brighter than on the Palomar
Sky Survey plates.  Contemporaneous 
optical spectroscopy acquired during this high state 
finds a weak emission line that we attribute to \ion{Mg}{2} $\lambda 2798$
at redshift $z = 0.75$, supporting a
flat spectrum radio quasar (FSRQ) classification.
The variability characteristics and radio properties together indicate that 
0FGL J1830.3+0617 at Galactic latitude $b = +7.\!^\circ5$
is a blazar.  Blazar identifications of three additional low-latitude
\fermi\ sources, 0FGL J0643.2+0858, 0FGL J1326.6$-$5302, and
0FGL J1328.8$-$5604, are also suggested.
\end{abstract}

\keywords{gamma rays: observations -- X-rays: individual (Swift J1830.1+0619)}

\section{Introduction}

The complete identification of samples of
high-energy ($> 200$ MeV) $\gamma$-rays
sources remains as one of the outstanding 
challenges in modern astrophysics. The main difficulty arises from the 
limited angular resolution (from a few arcminutes to
degrees) that can be achieved at GeV 
energies, which allows for multiple candidate counterparts within
the $\gamma$-ray error circles.  Overcoming the positional obstacles, 
the EGRET instrument \citep{thomp93} managed to identify
blazars and pulsars as the dominant contributors 
to the $\gamma$-ray source population 
\citep{hartman}. However, despite significant 
observational efforts, more than half of the sources catalogued
by EGRET remained unidentified \citep{thompson}. 

During its first months of operation, the {\it Fermi Gamma-ray Space 
Telescope\/} (\fermi) has 
improved the sensitivity and angular 
resolution achieved by EGRET, regaling the
high-energy community with an improved view of the 
GeV sky \citep{abdo1}. Of the 205 most 
significant sources reported 
in the Bright Gamma-ray Source List (0FGL), 168 have 
been associated with blazars, pulsars, high-mass X-ray binaries, 
radio galaxies, pulsar wind
nebulae, supernova remnants, a globular cluster, and the LMC. Among the
0FGL sources, 37 have no obvious counterparts at other wavelengths. 

Multiwavelength
observations play a powerful role
in the ultimate goal of producing definite 
identifications for 
all $\gamma$-ray emitters. 
In particular, the absence of obvious 
counterparts for 18\% of $\gamma$-ray sources 
in the 0FGL admits the possibility
that there are additional types of $\gamma$-ray emitters 
\citep{montmerle} or even potentially 
exotic phenomena \citep{baltz}. It is only through dedicated 
multiwavelength programs that the likelihood of new 
types of $\gamma$-ray emitters can be assessed.

Previous multiwavelength 
studies of $\gamma$-ray sources without obvious counterparts 
have yielded some noteworthy results.  For example,
\citet{mirabal} identified the X-ray counterpart of 3EG J1835+5918 
as the second member of a much larger population of radio-quiet pulsars
that has been recently confirmed by \fermi\ \citep{abdo1}. Multiwavelength
efforts also suggested the emergence of a class of $\gamma$-ray 
sources associated with 
radio galaxies \citep{muk2,combi}. 

Building on the improved localizations and sensitivity
achieved by {\it Fermi}, we have started a dedicated multiwavelength
program to investigate the nature of unidentified sources in the 0FGL.
In a parallel effort, 
\citet{bassani} suggested a probable association of
\fermi\ source 0FGL J2001.2+4352 with a BL Lac object.
In this Letter, we report the association of 
0FGL J1830.3+0617, a low Galactic latitude source
at $(\ell, b)=(36.\!^{\circ}158,+7.\!{^\circ}543)$,
listed as having photon flux $1.7 \times 10^{-7}$
cm$^{-2}$ s$^{-1}$ above 100~MeV,
with the flat-spectrum radio quasar (FSRQ)
B1827+0617.
The organization of this paper is as follows. In \S 2, we describe
multiwavelength observations of B1827+0617. 
In \S 3 we discuss the implications of our results. 
Finally, conclusions are presented in \S 4.

\section{Observations}

\begin{figure}[t]
\hfil
\includegraphics[width=1.\linewidth]{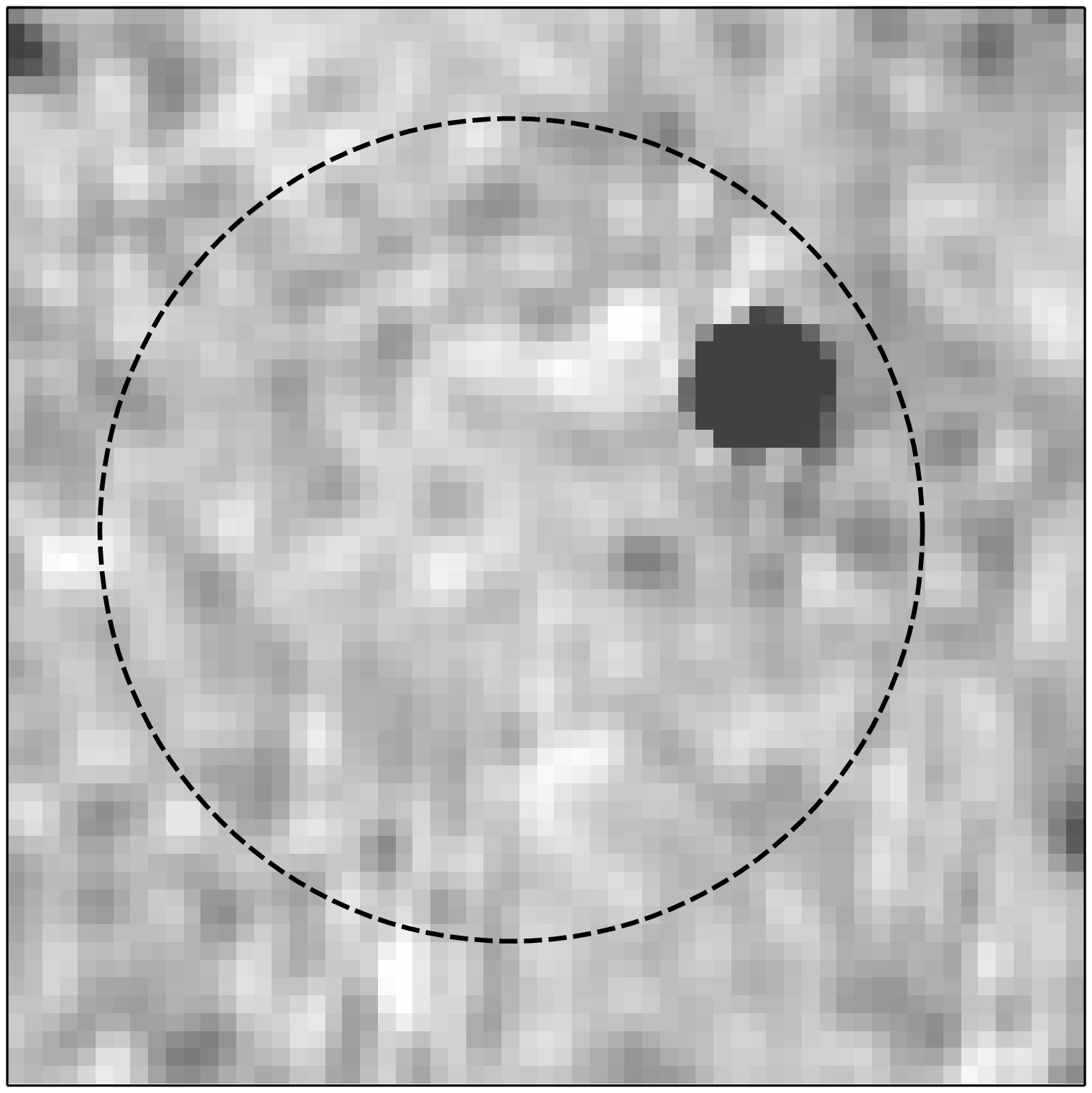}
\hfil
\caption{
NVSS postage stamp image of 0FGL J1830.3+0617 at 1.4 GHz
overlain with the $5.\arcmin8$ radius \fermi\ error
circle derived by \citet{abdo1}. The bright source corresponds to
NVSS J183005.90+061916.4 (B1827+0617). 
The field is $15\arcmin \times 15\arcmin$.
}
\label{radio}
\end{figure}

\begin{figure*}
\hfil
\includegraphics[width=7.5cm]{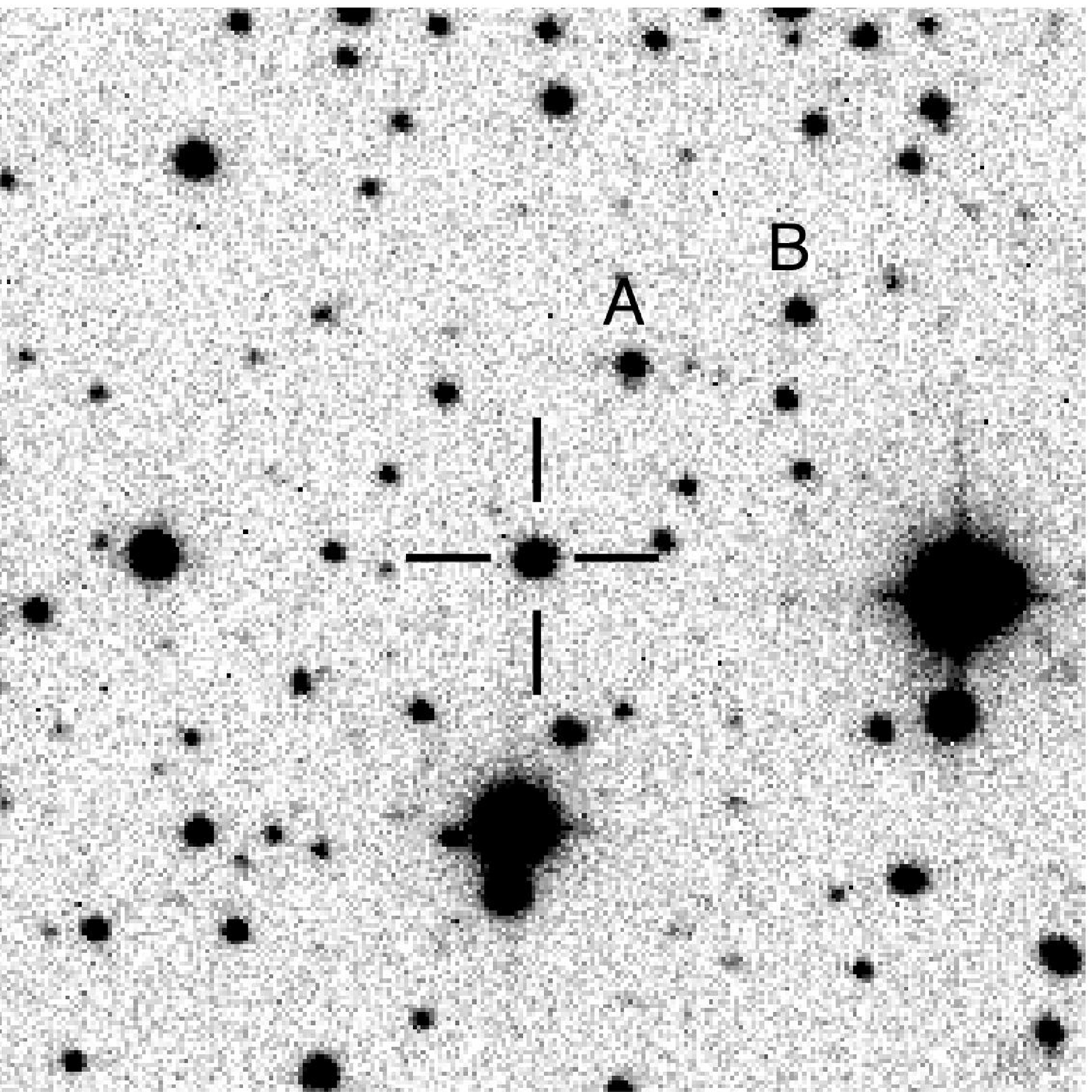}
\hfil
\includegraphics[width=7.5cm]{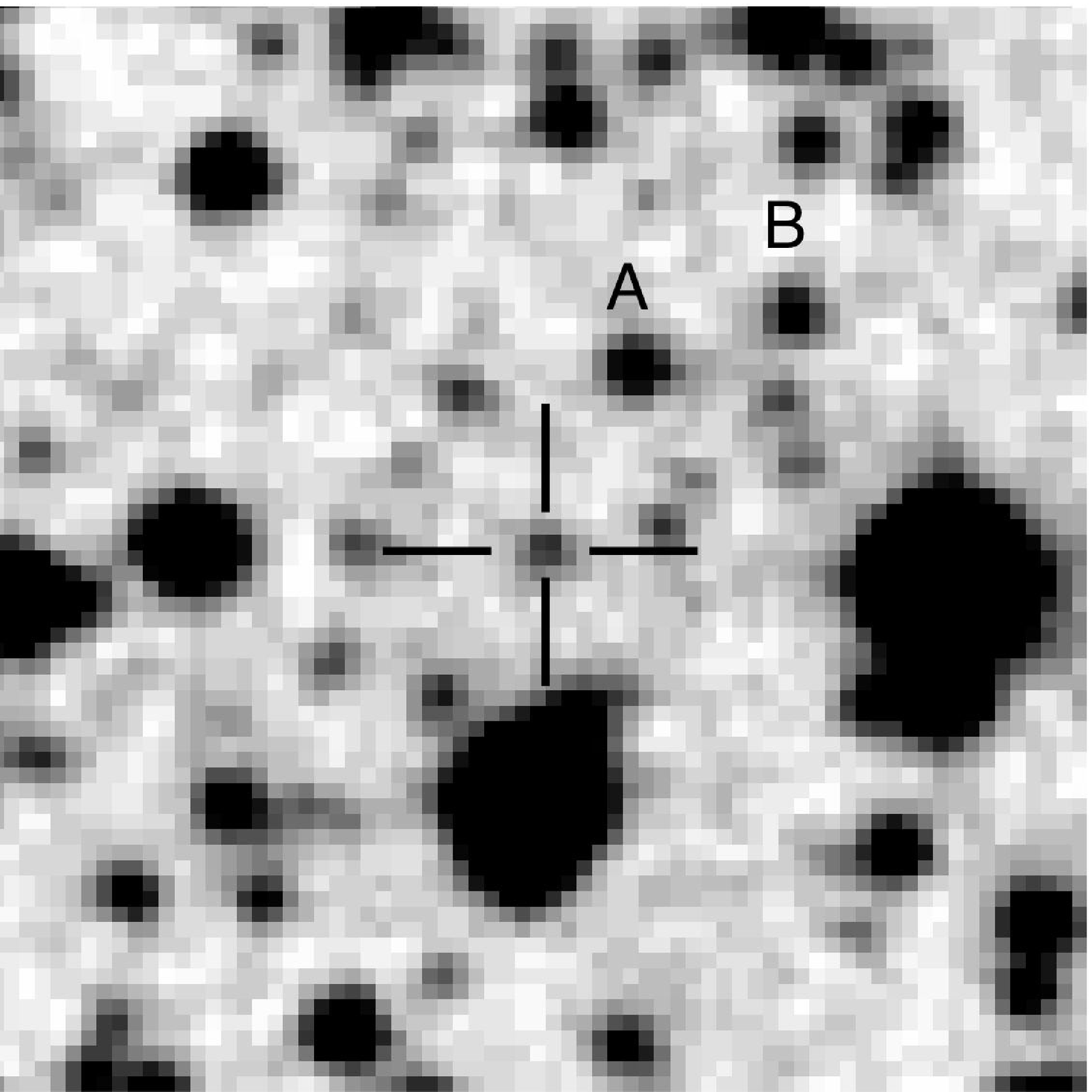}
\hfil
\caption{
{\em Left\/}: RETROCAM $r$-band image
obtained with the MDM 2.4 m telescope on
2009 May 29 UT at a scale of $0.\!^{\prime\prime}259$~pixel$^{-1}$.
The field is $70^{\prime\prime}$ across.
North is up, and east is to the left.
The optical counterpart of B1827+0617, at coordinates
(J2000.0) R.A.=$18^{\rm h}30^{\rm m}05.\!^{\rm s}95$,
decl.=$+06^{\circ}19\arcmin16\farcs2$,
is marked with a cross.
Comparison stars A and B have USNO-B1.0 magnitudes $R=18.4$ and
$R=18.8$, respectively, while B1827+0617 has $R \approx 16.9$.
{\em Right\/}: Digitized Palomar Sky Survey red plate
($1^{\prime\prime}$~pixel$^{-1}$)
of the same field from 1993 May 27 UT.
}
\label{photo}
\end{figure*}

\subsection{Radio}

A number of archival radio observations have covered the field of
0FGL J1830.3+0617. An examination of the 
NRAO VLA Sky Survey (NVSS) source catalog reveals  
only one source within the 95\% confidence level radius derived by 
\fermi.  NVSS J183005.90+061916.4 is listed as an unresolved
source of flux density 397 mJy at 1.4 GHz \citep{condon}. 
In order to provide a map of the entire field, 
we generated a radio finding chart from the 1995 February 27 VLA 
observation using the postage stamp image server available on the 
NVSS World Wide Web\footnote{Available at 
http://www.cv.nrao.edu/nvss/postage.shtml}. Figure~\ref{radio} 
shows the 1.4 GHz image and the localization of
NVSS J183005.90+061916.4 within the \fermi\ error circle.
A corresponding source GB6 J1830+0619 (B1827+0617 herein)
was also detected with a flux density of 443 mJy at 4.85 GHz
in the Green Bank 4.85 GHz northern sky survey carried out during
1986 November and 1987 October \citep{gregory}. 
We accordingly estimate a spectral index $\alpha = 0.09$ between 1.4 
and 4.85 GHz, defined as
$S_{\nu} \propto \nu^{\alpha}$. Given the resulting spectral index, 
we classify B1827+0617 as a flat-spectrum 
($\alpha > -0.5$) radio source.

\subsection{X-rays}
The X-Ray Telescope (XRT) on board the \swift\ observatory \citep{gehrels}
observed the field of 0FGL J1830.3+0617 
on 2009 May 20 UT for a total of 580~s of useful
exposure in photon counting (PC) mode. For the data reduction,
we used grades 0--12 and version 3.3 of the \swift\ software.  
As in the radio observations, a single source, 
Swift J1830.1+0619, was detected with {\it xrtcentroid} within the 
\fermi\ error circle at (J2000.0) R.A.=$18^{\rm h}30^{\rm m}05.\!^{\rm s}8$,
decl.=$+06^{\circ}19\arcmin12\arcsec$ with a 6\arcsec\ uncertainty, 
consistent with the position of 
B1827+0617. X-ray counts were extracted from a circular
region with a 20 pixel radius (47\arcsec). The background was
extracted from a source-free region with similar radius. The count
rate obtained from the observation is 
$(5.3 \pm 0.8) \times 10^{-2}$ s$^{-1}$. Although the
source contains only 31 photons, Swift J1830.1+0619 is detected at a  
5.5 $\sigma$ level of significance. Because of the limited number of
photons detected, no spectral fitting was attempted. Using 
WebPIMMS\footnote{http://heasarc.gsfc.nasa.gov/Tools/w3pimms.html}, we
estimate an absorbed flux in the 
0.3--10 keV band of $(2.2 \pm 0.3) \times 10^{-12}$
erg cm$^{-2}$ s$^{-1}$ assuming
a power law spectrum with $\Gamma = 2.0$ and a Galactic \ion{H}{1}
column density $N_{\rm H}$ = 
$2.4 \times 10^{21}$ cm$^{-2}$ as obtained from the 
nH tool\footnote{http://heasarc.gsfc.nasa.gov/cgi-bin/Tools/w3nh/w3nh.pl}. 
We note that $N_{\rm H}$
is consistent with Galactic absorption derived assuming an optical
extinction $E(B - V) = 0.46$ \citep{schlegel} and 
the standard conversion $N_{\rm H} / E(B - V)                        
= 5.0 \times 10^{21}$ cm$^{-2}$ mag$^{-1}$ \citep{savage}.

\subsection{Optical}

Optical identification of B1827+0617 was made
during 2009 May 25 and May 29 UT using the RETROCAM 
imager \citep{mor05} mounted on the MDM
2.4 m Hiltner Telescope. Images in Sloan $g,r,i,z$
were taken, and the $r$-band image is shown Figure~\ref{photo}.
Magnitudes were measured using  
a $1\farcs5$ radius aperture centered at the optical position 
of the object at coordinates
(J2000.0) R.A.=$18^{\rm h}30^{\rm m}05.\!^{\rm s}95$,
decl.=$+06^{\circ}19\arcmin16\farcs2$ in the 
USNO-B1.0 astrometric system.
An approximate $R$ magnitude was derived using
USNO magnitudes of nearby stars. 
In this system, stars A and B in Figure~\ref{photo}
have $R = 18.4$ and $R = 18.8$, respectively.

On both nights, B1827+0617 was found at
$R \approx 16.9$.  It is clear that this is
brighter by at least 2 magnitudes in comparison 
with any of the
Palomar Sky Survey plates obtained in 1950, 1990, and 1993.
Visual inspection shows that B1827+0617 was definitely
fainter than star B on the Digitized Sky Survey image from
the 1993 May 27 plate, also shown in Figure~\ref{photo}.
Therefore, reporting here a $>2$ mag high state in 2009
is rather conservative.

\begin{figure}[b]
\centerline{
\hfil
\includegraphics[width=0.95\linewidth]{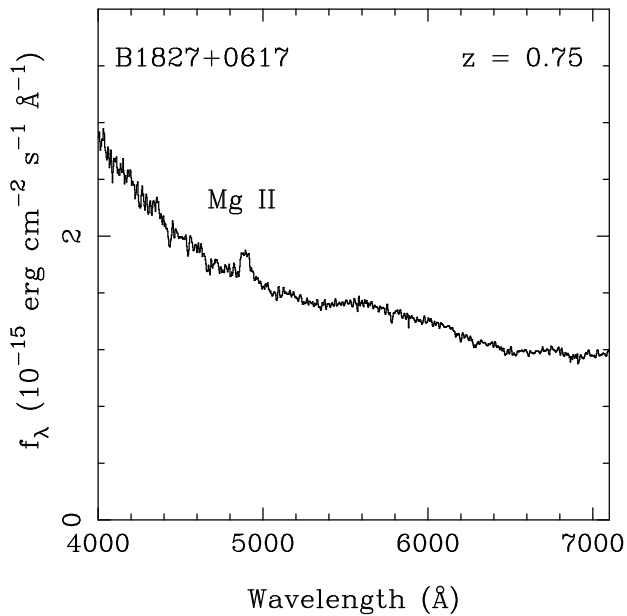}\hfill
\hfil
}
\caption{Spectrum of B1827+0617
obtained on 2009 May 25 UT, dereddened
assuming $E(B - V) = 0.46$ mag \citep{schlegel}. The single emission
line Mg II $\lambda2798$ at $z = 0.75$ is indicated.}
\label{spectrum}
\end{figure}

In addition to the optical imaging, we used
the Boller \& Chivens CCD spectrograph (CCDS) mounted on the 
MDM 2.4 m telescope to acquire spectra of B1827+0617.
The setup used provides 3.1 \AA~pixel$^{-1}$ dispersion and 
$\approx 8.0$~\AA~resolution with a $1\arcsec$ slit. 
Observations consisted of three 1200 s integrations 
obtained on 2009 May 25 UT under photometric conditions. 
The spectra were processed using standard procedures in IRAF\footnote{IRAF 
is distributed by the National Optical Astronomy Observatories,
    which are operated by the Association of Universities for Research
    in Astronomy, Inc., under cooperative agreement with the National
    Science Foundation.} and a final spectrum was generated by
combining the three individual exposures. 
The wavelength scale was established by fitting a set of
polynomials to Xe lamp spectra. A Hg-Ne lamp was also used 
to verify the wavelength calibration. Finally, we derived the 
flux calibration from observations of 
the spectrophotometric standard Feige 34 \citep{stone} observed at 
comparable telescope pointing to B1827+0617.

Galactic reddening is a significant factor
in this line of sight at Galactic coordinates 
$(\ell, b)=(36.\!^{\circ}158,+7.\!^{\circ}543)$. Therefore, we produced 
a final dereddened spectrum using an extinction correction 
of $E(B-V) = 0.46$ for these coordinates derived  
from the dust maps of \citet{schlegel}. 
Figure~\ref{spectrum} shows the resulting dereddened, 
wavelength- and flux-calibrated spectrum of B1827+0617.
A single weak emission line at 4892 \AA\ can be identified
with \ion{Mg}{2} $\lambda 2798$ at a redshift of $z = 0.75$ by
process of elimination, considering other expected
QSO emission lines that are not seen for alternative
identifications.  In addition, the weak, broad shelf
to the red of \ion{Mg}{2}, with a maximum at 5600 \AA,
is due to the blend of \ion{Fe}{2} multiplets that
commonly contribute to the feature known as the small blue bump
in QSO spectra.  As is often the case in blazars,
these and other emission lines may become
more prominent in the spectrum when the continuum
declines to a low state.

\section{Discussion}

A typical $\gamma$-ray error circle contains several radio and X-ray 
sources to contend with.
The field of 0FGL J1830.3+0617 is rather simple in that
a single radio/X-ray source, B1827+0617, emerges as the only 
counterpart candidate. 
Optical photometry shows that B1827+0617 is at least
2 magnitudes brighter than its historical level as seen
on sky survey plates.
Interestingly, there is also evidence for variability
of 0FGL J1830.3+0617 in \fermi\ observations
between 2008 August 4 and 2008 October 30
\citep{abdo1}. Multiwavelength variability and periods of enhanced activity are
typically observed in  $\gamma$-ray blazars 
\citep{abdo2}.

\begin{figure}[t]
\centerline{
\hfil                                                          
\includegraphics[width=0.95\linewidth]{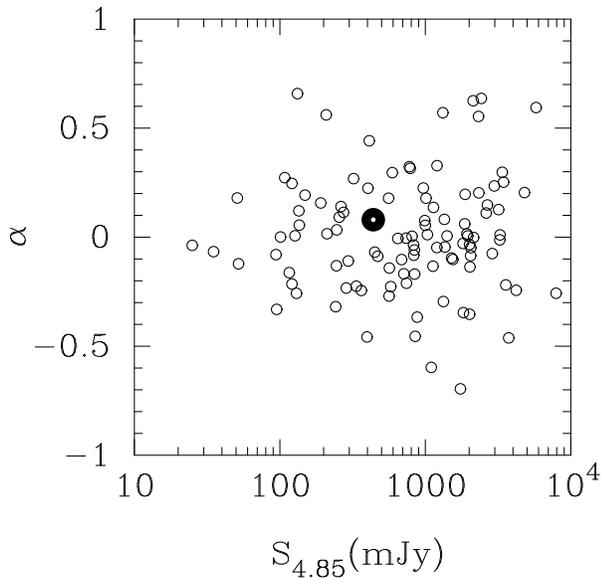}\hfill
\hfil
}
\caption{
Spectral index $\alpha$ between 1.4 and 4.85 GHz
as a function of flux density at 4.85 GHz S$_{4.85}$ 
for 100 blazars detected by \fermi\ \citep{abdo1}. B1827+0617 is
indicated by the large filled circle.}
\label{population}
\end{figure}

Turning our attention to the radio band, archival
measurements indicate a spectral index of 
$\alpha = 0.09$ between 1.4 and 4.85 GHz. In order to compare
the radio properties of B1827+0617 with proposed blazar associations 
in the 0FGL \citep{abdo1}, we culled measurements from the
Green Bank (GB6) catalog at 4.85 GHz \citep{gregory}, 
the 4.85 GHZ Parkes-MIT-NRAO (PMN) survey 
catalogue \citep{griffith}, and the 1.4 GHz NVSS catalog \citep{condon}.
For sources classified as blazars \citep{abdo1}, flux densities were
assembled and a spectral index $\alpha$ (where $S_{\nu} \propto \nu^{\alpha}$)
computed between 1.4 and 4.85 GHz. 
We show in Figure~\ref{population} 
the radio spectral index distribution as a function of the
flux density at 4.85 GHz for 100 \fermi\ blazars. 
Clearly, the results indicate that the radio properties of B1827+0617 are in
accord with the flux densities and flat spectral indices of
typical blazars.

The association of B1827+0617 with 0FGL J1830.3+0617 
is further supported by the optical 
spectrum of B1827+0617, which indicates a redshift $z = 0.75$. 
The presence of a single 
emission line in the optical spectrum
accompanied by a flat-spectrum radio spectral index
is consistent with a FSRQ blazar.

\begin{figure*}[t]
\hfil
\includegraphics[width=2.1in,angle=0.]{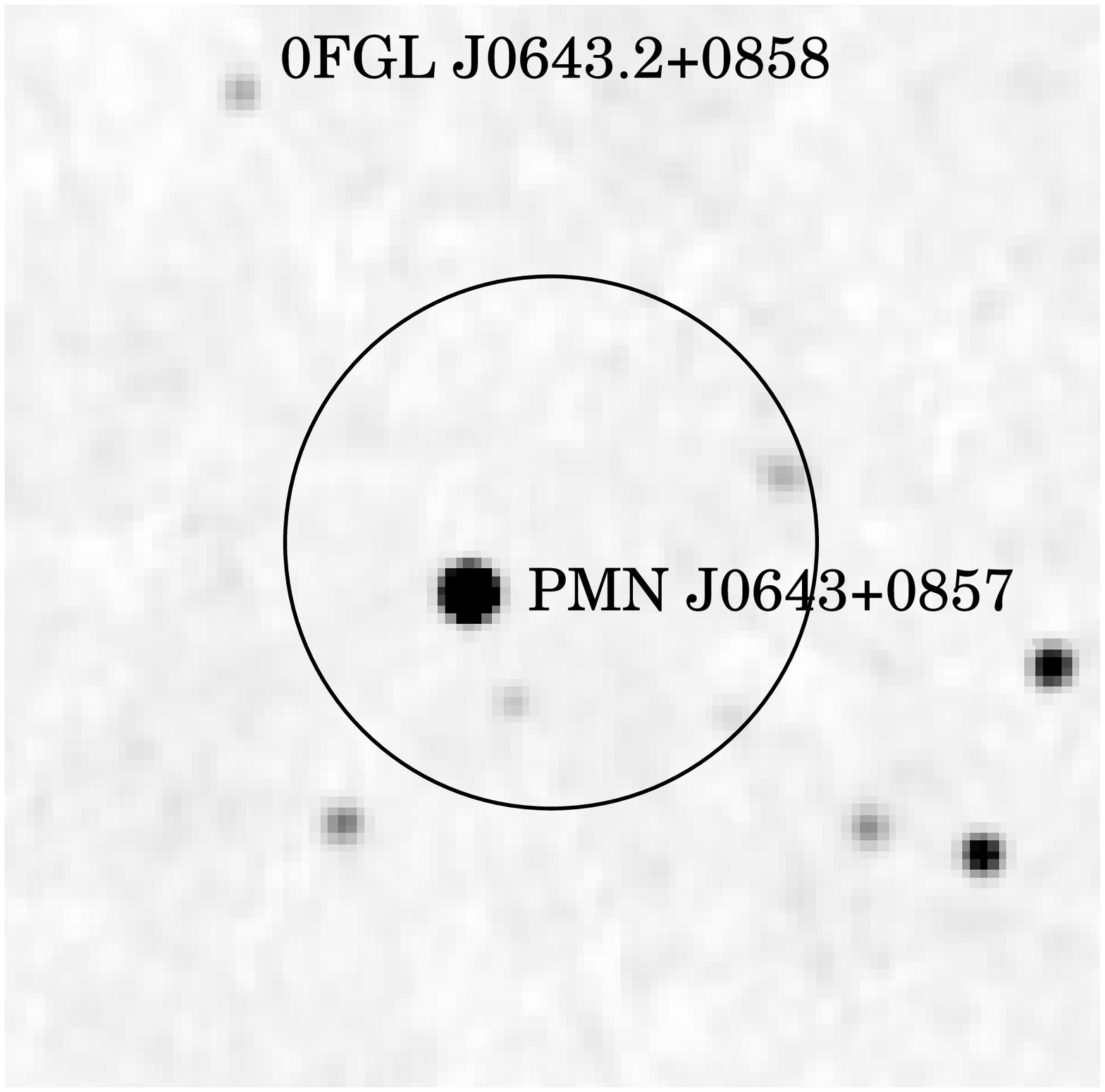}
\includegraphics[width=2.1in,angle=0.]{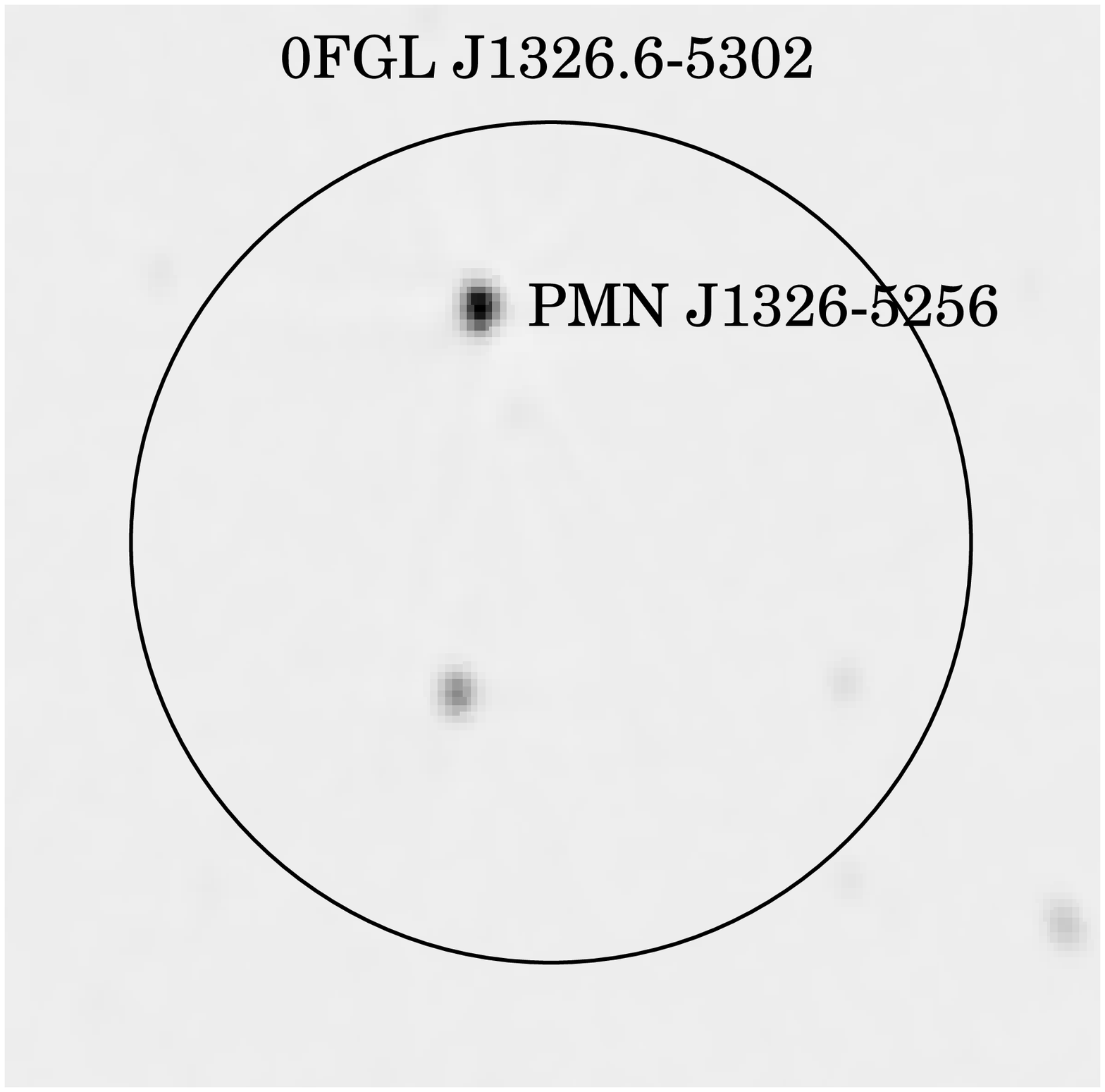}
\includegraphics[width=2.1in,angle=0.]{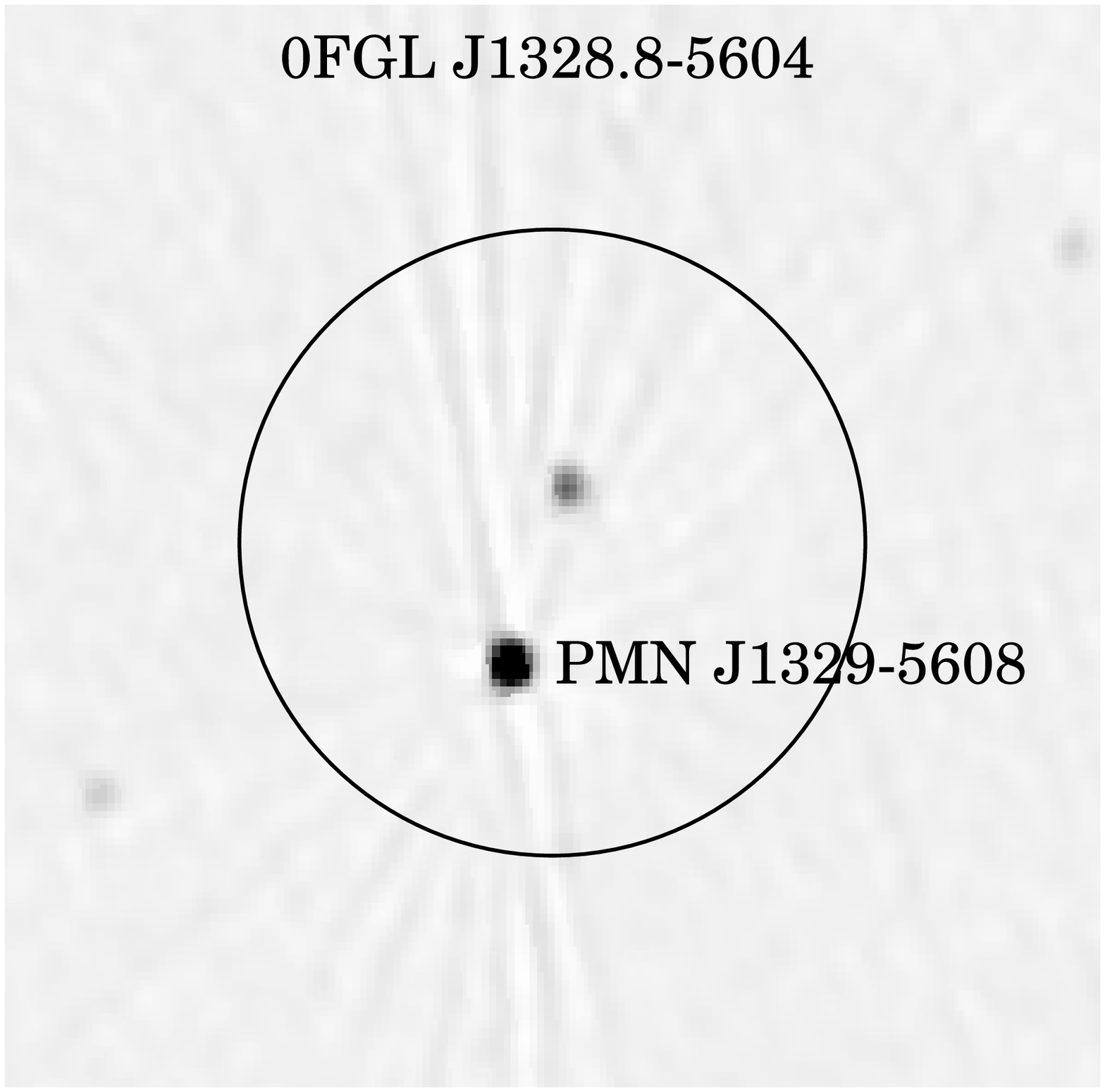}
\hfil
\caption{
Radio images of unidentified {\it Fermi} sources with error
circles from Abdo et al. (2009a).  Proposed flat-spectrum
radio source identifications are indicated (see text for
radio references).  Each image is $0.\!^{\circ}5$ across.
0FGL J0643.2+0858 is from the 1.4~GHz NVSS, while the two
southern sources are from the 843 MHz Sydney University Molonglo
Sky Survey \citep{bock}
}
\label{figure5}
\end{figure*}

Statistically, as many blazars are expected
to be located near the Galactic plane as anywhere
else \citep{muk1,halpern,sguera}. 
Indeed, by extrapolation, \fermi\ should detect 20$-$25 
blazars at $|b|\leq 10^\circ$ \citep{abdo2}.  Thus far, 
five blazar associations are reported in the 0FGL 
at $|b| \leq10^\circ$. In addition,
0FGL J0910.2$-$5044 at $b = -1.\!^{\circ}8$
\citep{che08,lan08} is likely associated with
a blazar \citep{sadler}. 
The new identification of 
0FGL J1830.3+0617 confirms the expectation of 
additional \fermi\ blazars within the zone $-10^\circ < b < 10^\circ$. 
A blazar classification for B1827+0617 using the
BZCAT \citep{massaro}, 
CRATES \citep{healey1}, or
CGRaBS \citep{healey2} catalogs was most likely missed
because such surveys only aim for uniform sky coverage at
$|b|\geq 10^\circ$.   We anticipate that several additional
``unidentified'' \fermi\ sources at low Galactic latitude
can be associated with known flat-spectrum radio sources.
Among our proposed identifications (Figure~\ref{figure5}) are
0FGL J0643.2+0858 = PMN J0643+0857 \citep{gri95,pet06},
0FGL J1326.6$-$5302 = PMN J1326$-$5256 and
0FGL J1328.8$-$5604 =  PMN J1329$-$5608 \citep{gri95,mas08}.

\section{Conclusions}

We have presented radio, optical, and X-ray observations of 
the field of the
unidentified \fermi\ source 0FGL J1830.3+0617.
Only a single plausible counterpart, B1827+0617, 
is detected within its 95\% error circle. In the absence of
other alternatives, we have argued that 
the B1827+0617 is most likely associated with 
0FGL J1830.3+0617. A blazar identification of the \fermi\ source is supported 
by its flat spectral index between 1.4 and 4.85 GHz, as well as 
the detection of optical/$\gamma$-ray variability. The optical spectrum 
of B1827+0617 has a single emission line that we identify with
\ion{Mg}{2} $\lambda2798$ at $z = 0.75$, which supports 
a FSRQ blazar classification for this source. 
The radio properties of B1827+0617 are compatible with
known \fermi\ blazars in the 0FGL. These findings suggest that the proposed
blazar association is largely secure.  Additional identifications
of \fermi\ blazars at low Galactic latitude with known radio sources
can proceed in the same manner as for high-latitude sources.
Ultimately, the unequivocal identification of $\gamma$-ray
blazars will depend strongly on the detection of
contemporaneous variability in $\gamma$ rays and at
least one additional energy band.

Independent of the $\gamma$-ray strategies, it is crucial to move ahead with 
dedicated multiwavelength programs of both identified and 
unidentified \fermi\ sources. These efforts may
prove successful in sorting out 
the presence of novel classes of $\gamma$-ray emitters.
Such multiwavelength studies 
will increase in complexity as the \fermi\ mission continues to 
unveil the full distribution of GeV sources in the sky to even 
fainter flux levels.

\acknowledgments
NM acknowledges support from the Spanish Ministry of Science
and Technology through a Ram\'on y Cajal fellowship.

\end{document}